\documentclass[12pt]{article}
\usepackage[centertags]{amsmath}
\usepackage{amsfonts}
\usepackage{amssymb}
\usepackage{amsthm}
\usepackage{graphicx}
\usepackage{axodraw4j}
\newcommand{\beq}{\begin{equation}}
\newcommand{\eeq}[1]{\label{#1}\end{equation}}
\newcommand{\bea}{\begin{eqnarray}}
\newcommand{\eea}[1]{\label{#1}\end{eqnarray}}
\begin{document}

\title{
\vspace{1cm}
\Large\textbf{Consequences of Gravity-Induced Couplings in Theories
with Many Particle Species}
\vspace*{.5cm}
\author{
{\large\textbf{Rakibur Rahman\footnote{email: mrr290@nyu.edu} }}\\
\small \emph{Center for Cosmology and Particle Physics}\\
\small \emph{Department of Physics, New York University}\\
\small \emph{4 Washington Place, New York, NY 10003}}}

\date{}

\maketitle \thispagestyle{empty} \vspace*{.5cm}

\begin{abstract}
In theories with many copies of the Standard Model virtual black
hole exchange may produce effective higher dimensional operators
that can be treated below the cutoff scale as fundamental vertices
of interspecies non-gravitational interaction. We consider the
vertex that couples fermions of one species through magnetic moment
to photons of other species, and study the quantum corrections it
generates. In particular, we find kinetic mixing between photons of
different species produced via fermion loops. Diagonalization of
gauge kinetic terms then renders the fermions millicharged under
other species' electromagnetism. We explore some phenomenological
consequences of such effects by considering possible observable
signatures in collider experiments and constraining the interaction
strength. The derived bounds are in agreement with non-democratic
nature of micro black hole coupling.
\end{abstract}

\newpage
\renewcommand{\thepage}{\arabic{page}}
\setcounter{page}{1}

\section{Introduction}

Theories with a large number $N$ of particle species~\cite{d1}
provide a simple solution to the hierarchy problem by lowering the
effective gravitational cutoff to a scale~\cite{dr}: \beq
\Lambda_G=\frac{M_P}{\sqrt{N}}\,.\eeq{i1} Lowering of the
gravitational cutoff finds justification from consistency of large
distance black hole physics~\cite{d1,dr}, from black hole entropy
considerations~\cite{ds}, and from quantum information
theory~\cite{dg}. The hierarchy problem is explained by the large
$N$ number of species with cutoff scale $\Lambda$ by assuming that
gravity becomes strong at the scale $\Lambda_G\approx\Lambda$. Given
this one will start probing quantum gravity effects in particle
collisions around the scale $\Lambda\sim\text{TeV}$. In particular,
production of microscopic black holes will take place by collisions
of Standard Model (SM) particles at that energy scale.

In this scenario virtual black hole exchange can generate
interspecies scattering processes. The intermediate black hole
states are heavy, with mass $\mathcal{O}(\Lambda)$. They can be
integrated out below the scale $\Lambda$. By doing so one ends up
having effective higher dimensional operators, which of course will
be compatible with the symmetries. In the framework where the other
species are exact copies of the SM, or some other copies with
photon-like fields, one can have the following gauge invariant
operator: \beq \mathcal{L}_{\text{int}}=\sum_{i\neq j}
\frac{\lambda_{ij}}{M_\text{P}}\,F^{\mu\nu}_i\bar{\psi}_j
\sigma_{\mu\nu}\psi_j.\eeq{i2} Here fermions belonging to the $j$-th
species couple via magnetic moment to the photon in the $i$-th
species. The dimensionless coupling constants $\lambda_{ij}$ are
real, non-vanishing and non-diagonal. Below the scale $\Lambda$, the
above interaction can be treated as a fundamental vertex of
non-gravitational coupling among different particle species. Such
non-gravitational couplings may have very interesting consequences,
which we investigate in this paper. The point is that from fermion
loops there will be quantum corrections to the photon propagators,
which will render the photon kinetic terms non-diagonal\footnote{Here
one may argue that photon kinetic mixing terms could already be present
at tree level via dimension-4 operators. However, as our philosophy is to
consider different copies of the SM that are coupled \emph{only} gravitationally,
we have no such terms at tree level. They only arise from \emph{effective}
non-gravitational interactions produced by virtual black hole exchange.}.
One can always diagonalize the kinetic terms by redefining the photon
fields. But the latter inevitably endows fermions belonging to our SM
with tiny charges under electromagnetic gauge groups of other species.

The organization of this paper is as follows. In Section 2 we
compute Feynman diagrams involving fermion loops that renormalize
the photon propagators, and give rise to kinetic mixing between
photons of different species. The diagram is regularized by using
Pauli-Villars regularization technique~\cite{pv}. We then
diagonalize the photon kinetic terms, which results in redefinition
of fermion covariant derivatives. In particular, fermions become
millicharged under other species' electromagnetism. In Section 3 we
discuss phenomenological consequences of such effects by considering
scattering processes where we have an incoming fermion-antifermion
pair belonging to our SM, and an outgoing pair of other species.
Such processes, potentially observable in collider experiments, put
bounds on the dipole moment couplings. The draw concluding remarks
in Section 4.

\section{Photon Kinetic Mixing via Fermion Loop}

The effective interaction under consideration is: \beq
\mathcal{L}_{\text{int}}=\sum_{i\neq j}\frac{\lambda_{ij}}
{M_\text{P}}\,F^{\mu\nu}_i\bar{\psi}_j\sigma_{\mu\nu}\psi_j.\eeq{r1}
To find the momentum space Feynman rules, we define photon momentum
as outgoing, so that $\partial^\mu A^\nu\rightarrow
+ip^\mu\tilde{A}^\nu$. We have a vertex with an incoming
fermion-antifermion pair of the $j$-th species, and an outgoing
photon in the $i$-th species. If the photon carries a Lorentz index
$\nu$, with momentum $p^\mu$ attached to it, the corresponding
vertex-factor is: \beq \text{Vertex}=\frac{2i}{M_\text{P}}\,
\lambda_{ij}p^\mu\sigma_{\mu\nu}.\eeq{r2} Such a vertex will give
rise to diagrams where a photon $A_{k,\mu}$ of the species $k$ goes
into a photon $A_{i,\nu}$ of another species $i$ via fermion loop.\\
      \begin{center}
      \begin{picture}(200,25) (0,0)
       \Arc[arrow,arrowlength=-5,arrowwidth=2](80,0)(20,90,90)
       \ArrowLine(20,10)(35,10)\Text(27.5,17)[]{$p$}\Text(82,29)[]{$k,j$}
       \Text(27.5,-17)[]{$\mu,k$}\Text(132.5,-17)[]{$\nu,i$}
       \ArrowLine(125,10)(140,10)\Text(132.5,17)[]{$p$}\Text(82,-28)[]{$k-p,j$}      \Arc[arrow,arrowlength=-5,arrowwidth=2](80,0)(20,270,270)
       \Photon(100,0)(140,0){4}{4.5}
       \Photon(20,0)(60,0){4}{4.5}\Text(80,-47)[]{Figure 1}
     \end{picture}
     \end{center}
\vspace{2cm}
The graph is given by \bea \Pi^{\mu\nu}&=&
\left(\frac{2i}{M_\text{P}}\right)^2(-1)\int\frac{d^4k} {(2\pi)^4}
\sum_j\text{Tr}\left[\lambda_{ij}p_\rho\sigma^{\rho\nu}\left(\frac{i}
{\not{\!k\!}-m_j}\right)\lambda_{kj}(-p_\lambda)\sigma^{\lambda\mu}
\left(\frac{i}{\not{\!k\!-\not{\!p\!}-m_j}}\right)\right]\nonumber\\
\nonumber\\&=&\sum_j \lambda_{ij}\lambda_{kj}\Pi^{\mu\nu}_j(p,
m_j),\eea{r4} where $\Pi^{\mu\nu}_j(p,m_j)$ is defined as \beq
\Pi^{\mu\nu}_j (p,m_j)\equiv-\frac{4}{M_\text{P}^2}\,p_\rho
p_\lambda\int\frac{d^4k}{(2\pi)^4}\,\frac{k_\alpha(k-p)_\beta\,
\text{Tr}(\gamma^{\rho\nu}\gamma^\alpha\gamma^{\lambda\mu}\gamma^\beta)
+m_j^2\,\text{Tr}(\gamma^{\rho\nu}\gamma^{\lambda\mu})}{(k^2-m_j^2)
\{(k-p)^2-m_j^2\}}~.\eeq{r5}

The above integral seems quadratically divergent for large internal
momentum $k$. To give it a meaning we employ the technique of
Pauli-Villars regularization~\cite{pv}. This amounts to minimally
coupling the photons to additional spinor fields with a very large
mass $M_s$. These fields might have ghost couplings. This
prescription implies the replacement~\cite{pv}: \beq
\Pi^{\mu\nu}_j(p,m_j) \rightarrow \bar{\Pi}^{\mu\nu}_j(p)
=\Pi^{\mu\nu}_j(p,m_j)+ \sum_{s=1}^SC_s \Pi^{\mu\nu}_j(p,M_s),
\eeq{r6} where the constants $C_s$ will be chosen such that the
integrals converge. The minimal coupling of the additional fields
implies that gauge invariance is preserved by the regularization
procedure.

A convenient way to evaluate~(\ref{r6}) is to introduce auxiliary
variables to elevate the propagator denominators into exponential
factors by the identity: \beq \frac{i}{k^2-m^2+i\epsilon}=
\int_0^\infty d\alpha\,e^{i\alpha(k^2-m^2+i\epsilon)}.\eeq{r7} After
some shift of variable one can perform the momentum integrals to
obtain \bea \Pi^{\mu\nu}_j(p,m_j)&=&\frac{-i} {\pi^2M_p^2}\,(p^\mu
p^\nu-\eta^{\mu\nu}p^2)\int_0^{\infty}\int_0^{\infty}\frac{d\alpha_1d
\alpha_2}{(\alpha_1+\alpha_2)^2}\,e^{-i(\alpha_1+\alpha_2)\left\{m_j^2
-\frac{\alpha_1\alpha_2}{(\alpha_1+\alpha_2)^2}\,p^2\right\}}\nonumber\\
\nonumber\\&&~~~~~~~~~~~~~~~~~~~~~~~~~~~~~~~~~~~~\times\left[\,\frac{\alpha_1
\alpha_2}{(\alpha_1+\alpha_2)^2}\,p^2+m_j^2\,\right].\eea{r14}
Notice that $\Pi^{\mu\nu}_j(p,m_j)$ is proportional to $(p^\mu
p^\nu-\eta^{\mu\nu}p^2)$, as required by gauge invariance. To
perform the integrations in~(\ref{r14}), we use the identity: \beq
1=\int_0^\infty \frac{d\lambda}{\lambda}\,\delta\left(1-
\frac{\alpha_1+\alpha_2}{\lambda}\right).\eeq{r15} Inserting this
into~(\ref{r14}), and then making the rescaling $\alpha_i\rightarrow
\lambda\alpha_i$, we obtain \bea \Pi^{\mu\nu}_j(p,m_j)&=&\frac{-i}
{\pi^2M_p^2}\,(p^\mu p^\nu-\eta^{\mu\nu}p^2)\int_0^{\infty}
\int_0^{\infty}d\alpha_1d\alpha_2\,\delta(1-\alpha_1-\alpha_2)
(\alpha_1\alpha_2p^2+m_j^2)\nonumber\\\nonumber\\
&&~~~~~~~~~~~~~~~~~~~~~~~~~~~~~~~~~~~~~~~\times\int_0^\infty
\frac{d\lambda}{\lambda}\,e^{-i\lambda (m_j^2-\alpha_1\alpha_2p^2)}.
\eea{r16} The integral over $\lambda$ looks logarithmically
divergent at the point $\lambda=0$. As we will see, if we choose the
coefficients $C_s$ in~(\ref{r6}) such that \bea C_s&=&-2\delta^1_s+
\delta^2_s,\label{r17}\\\nonumber\\\sum_{s=1}^S C_sM_s^2 &=&-2M_1^2+
M_2^2=-m_j^2,\eea{r18} we can render~(\ref{r16}) convergent.
Furthermore, let us pick $p$ such that $p^2<4m_j^2$, which
corresponds to the threshold of pair creation. Because $\alpha_1$,
$\alpha_2$ are positive definite and satisfy $\alpha_1+\alpha_2 =1$,
we have $\alpha_1\alpha_2\leq1/4$. Therefore the quantity
($m_j^2-\alpha_1\alpha_2p^2$) is positive and the integration
contour in the complex $\lambda$ plane can be rotated by $-\pi/2$,
so that \beq \int_0^\infty\frac{d\lambda}{\lambda}\,
e^{-i\lambda(m_j^2-\alpha_1\alpha_2p^2)} ~\rightarrow~
\ln\left(\frac{\Lambda^2}{m_j^2}\right)-\ln\left(1-\frac{\alpha_1
\alpha_2p^2}{m_j^2}\right),\eeq{r19} where we have defined \beq
\Lambda\equiv\frac{M_1^2}{M_2}~.\eeq{r20} The above combined with
condition~(\ref{r18}) gives \bea a_1&\equiv&
\frac{M_1^2}{\Lambda^2}=2\left(1-\frac{m_j^2}{4\Lambda^2}\right)
+\mathcal{O} (m_j^4/\Lambda^4),\label{r21}\\\nonumber\\
a_2&\equiv& \frac{M_2^2}{\Lambda^2}=4\left(1-\frac{m_j^2}
{2\Lambda^2}\right)+\mathcal{O}(m_j^4/\Lambda^4).\eea{r22} The
choice~(\ref{r17},\ref{r18}) also gives us \bea m_j^2&\times&
\int_0^\infty\frac{d\lambda}{\lambda} e^{-i\lambda(m_j^2-\alpha_1
\alpha_2 p^2)}\nonumber\\&\rightarrow& m_j^2\left[\,\ln\left(\frac
{\Lambda^2}{m_j^2}\right)-\ln\left(1-\frac{\alpha_1\alpha_2p^2}{m_j^2}
\right)\right]-4\ln2\,\Lambda^2+(3\ln2+1)\,m_j^2, \eea{r24} where we
have used the definitions of $a_1$, $a_2$, and some rescalings.
Plugging the regularized integrals~(\ref{r19},\ref{r24})
into~(\ref{r16}), we finally obtain \bea \bar{\Pi}^{\mu\nu}_j(p)
&=&\frac{i}{\pi^2M_p^2}\,(p^\mu p^\nu-\eta^{\mu \nu}p^2)\int_0^1
d\alpha\,[\,\alpha(1-\alpha)p^2 +m_j^2\,]\,\ln\left[1-\frac{
\alpha(1-\alpha)p^2}{m_j^2}\right] \nonumber\\\nonumber\\&&
-\,\frac{i}{6\pi^2M_p^2}\,\ln\left(\frac {\Lambda^2}{m_j^2}\right)
(p^2+6m_j^2)\,(p^\mu p^\nu-\eta^{\mu\nu} p^2)\nonumber\\\nonumber\\
&&+\,\frac{i}{\pi^2M_p^2}\,[\,4\ln2\, \Lambda^2-(3\ln2+1)\,
m_j^2\,]\,(p^\mu p^\nu-\eta^{\mu\nu}p^2).\eea{r25} In the limit
$\Lambda\rightarrow \infty$, then the graph~(\ref{r4}) reduces to
\beq \Pi^{\mu\nu}=\left(\frac{4i\ln2} {\pi^2}\right)\left(\frac{
\Lambda^2}{M_p^2}\right)\sum_j\lambda_{ij} \lambda_{kj}\,(p^\mu
p^\nu-\eta^{\mu\nu}p^2).\eeq{r26} Note that higher derivatives do
not appear in the leading divergent terms. 
The above radiative correction renormalizes the photon
propagators as: \beq -\frac{i\eta_{\mu\nu}}{p^2}\,\delta_{ij}
~\rightarrow~-\frac{i\eta_{\mu\nu}}{p^2}\,\left[\,\delta_{ij}
-\left(\frac{4\ln2}{\pi^2}\right)\,\frac{ \Lambda^2}{M_p^2}\,
\sum_k\lambda_{ik} \lambda_{jk}\,\right].\eeq{r27}

We need to redefine the photon fields in order to remove their
kinetic mixing, i.e. to obtain canonical kinetic terms. Let us
assume that all the species are strictly identical, i.e. exact
copies of the SM, related by certain permutation symmetry. Then all
the non-diagonal elements of the coupling matrix $\lambda_{ij}$ are
the same: $\lambda_{ij}=\lambda$, for $i\neq j$, while
$\lambda_{ii}=0$. This gives \beq \Delta_{ij}~\equiv~\left(\frac{
4\ln2}{\pi^2}\right)\,\frac{ \Lambda^2}{M_p^2}\,\sum_k\lambda_{ik}
\lambda_{jk}~\approx~\delta\left(\begin{array}{cccc}
                           1 & 1 & ... & 1 \\
                           1 & 1 & ... & 1 \\
                           ... & ... & ... & ... \\
                           1 & 1 & ... & 1
                         \end{array}\right),\eeq{r28}
where \beq \delta~\equiv~\left(\frac{4\ln2}{\pi^2}\right)\,\frac{
\Lambda^2}{M_p^2}\,(N-1)\lambda^2~\approx~\left(\frac{4\ln2}{\pi^2}
\right)\lambda^2~\ll~1.\eeq{r29} Because the matrix $\Delta_{ij}$ is
symmetric, the photon kinetic terms can be diagonalized by the field
redefinitions: \beq A_i^\mu\rightarrow \sum_j \left(\delta_{ij}-
\tfrac{1}{2}\Delta_{ij}\right)A_j^\mu. \eeq{r30} This will result in
redefinition of fermion covariant derivatives, so that fermions will
acquire millicharge under other species' electromagnetism: \beq
\mathcal{L}_{\text{Dirac}}=-i\sum_{i=1}^N\bar{\psi}_i\left[\,
\not{\!\partial\!}-m_i-ie\not{\!\!A}_i+ie(\delta/2)\sum_{j=1}^N
\epsilon_{ij}\not{\!\!A}_j\,\right]\,\psi_i, \eeq{r31} where
$\epsilon_{ij}=1$, for $i\neq j$, while $\epsilon_{ii}=0$.

\section{Phenomenological Consequences}

As long as our Standard Model is concerned, there are two kinds of
vertices given by the Lagrangian~(\ref{r31}): the one of usual QED,
and another QED-like vertex where the photon belongs to
\emph{another} SM. The latter vertex carries a tiny fermion charge
of $(-\delta/2)e$, instead of $e$. However, because of the enormous
multiplicity of species, the fermion millicharge may produce
observable signals in particle colliders.

Let us consider ultra-relativistic electron-positron
collision\footnote{The center of mass energy however must be taken
below the threshold of $t\bar{t}$ production. This is necessary for
the regularization of the loop integral, because otherwise the
condition $p^2<4m^2$ is never satisfied.}. At tree level the
$e^+e^-$ pair may produce a fermion-antifermion pair of \emph{other}
species either through the SM photon, or through other species'
photons. The corresponding Feynman diagrams are: \vspace{2cm}
\begin{center}
\begin{picture}(170,25) (0,0)
 \ArrowLine(0,30)(-30,60)\ArrowLine(30,60)(0,30)
 \ArrowLine(-30,-60)(0,-30)\ArrowLine(0,-30)(30,-60)
\Text(-35,-65)[]{$e^-$}\Text(37,-65)[]{$e^+$}\Text(-35,65)[]{$\psi^-_j$}\Text(37,67)[]{$\psi^+_j$}
\Text(-24,10)[]{Our}\Text(-24,0)[]{Photon}\Text(13,3)[]{$A_{\mu}$}
 \Photon(0,30)(0,-30){4}{4.5}
 \Vertex(0,30){1.5} \Vertex(0,-30){1.5}
\Text(40,30)[]{$-ie\gamma^{\mu}\left(-\delta/2\right)$}
\Text(20,-30)[]{$-ie\gamma^{\mu}$}
\Text(0,-80)[]{Figure 2}
 \end{picture}
\begin{picture}(20,25) (0,0)
 \ArrowLine(0,30)(-30,60)\ArrowLine(30,60)(0,30)
 \ArrowLine(-30,-60)(0,-30)\ArrowLine(0,-30)(30,-60)
 \Text(-35,65)[]{$\psi^-_j$}\Text(37,67)[]{$\psi^+_j$}
\Text(-35,-65)[]{$e^-$}\Text(37,-65)[]{$e^+$}
\Text(-24,10)[]{Other}\Text(-24,0)[]{Photon}\Text(13,3)[]{$A_{\mu i}$}
 \Photon(0,30)(0,-30){4}{4.5}
 \Vertex(0,30){1.5} \Vertex(0,-30){1.5}
\Text(40,-30)[]{$-ie\gamma^{\mu}\left(-\delta/2\right)$}
\Text(70,30)[]{$-ie\gamma^{\mu}\left(\delta_{ij}+\left(\delta_{ij}-1\right)\delta/2\right)$}
\Text(0,-80)[]{Figure 3}
 \end{picture}
\end{center}
\vspace{3cm}
The amplitude for the process, considering exact replicas of the SM, is given by:
\beq \mathcal{A}\,(e^+e^-\rightarrow\psi^+_j\psi^-_j)~=~-\tfrac{1}{2}\,\delta\mathcal{M}
-\tfrac{1}{2}\,\delta \left[\,1-\tfrac{1}{2}\,\delta(N-2)\,\right]\mathcal{M},\eeq{p1}
where $\mathcal{M}$ is the would-be amplitude if the outgoing pair belonged to our SM
instead. The process has a multiplicity of ($N-1$), and looks like $e^+e^-\rightarrow
\text{invisible}$, when viewed from our SM. The branching ratio is given by \beq \text{Br}\,
(e^+e^-\rightarrow\text{invisible})~=~(N-1)\,\frac{|\mathcal{A}|^2}{|\mathcal{M}|^2}
~\approx~\delta^2N\left(\tfrac{1}{4}\delta N-1\right)^2.\eeq{p2} Thus such processes
are potentially observable, because a small value of $\delta$ may be compensated by
a large value of $N$.

Alternatively, one can put bounds on the dipole moment coupling $\lambda$, by using~(\ref{p2}).
A stringent bound comes from invisible decays of orthopositronium~\cite{ortho}: \beq \text{Br}\,
(e^+e^- \rightarrow\text{invisible}) ~\leq~2.1\times 10^{-8}~(90\%\, \text{C.L.})\eeq{p3} This
gives, from~(\ref{r29}), with $N\sim 10^{32}$ copies: \beq \lambda~\lesssim~10^{-13}~\approx ~10^{3}
(\Lambda/M_P).\eeq{p4} We see that $\lambda$ has a suppression factor of $\Lambda/M_P$. This is in
fact compatible with the non-democratic nature of black hole couplings with different species~\cite{dr,d2},
which can be derived from unitarity considerations. Indeed, as shown in~\cite{dr}, the off-diagonal
couplings of the black holes must be suppressed at least by $1/\sqrt{N}$. After integrating them out
one finds that effective interspecies coupling is suppressed by the scale $\Lambda/M_P^2$.

\section{Conclusion}

In this paper we argued how in a theory of gravitationally coupled
many particle species effective non-gravitational interspecies
interactions arise because of virtual black hole exchange. We have
shown that through radiative corrections magnetic moment-type
interspecies coupling renders the SM fermions millicharged under
hidden sectors' electromagnetism. The effect may manifest itself in
particle colliders. By considering ultra-relativistic $e^+e^-$
collisions, we put bounds on such couplings, which are in agreement
with non-democratic nature of micro black hole coupling.

It is worth pointing out that many extensions of the SM, in
particular those coming from string theory, predict hidden $U(1)$
gauge groups, and naturally give rise to the kinetic mixing
phenomenon~\cite{kinmixing,abel}. According to the common lore,
gauge kinetic mixing is generated by irrelevant operators that
\emph{do} require the existence of cross-charged fundamental states
(e.g.~\cite{abel,hol}). However, such states are not indispensable.
Indeed, the magnetic moment interaction operator~(\ref{i2})
considered in this paper produces kinetic mixing without appealing
to cross-charged particles.

One would like to see what could be the implications of effective
non-gravitational interspecies couplings in the early universe
cosmology. Such considerations will probably put more severe bounds
on the coupling strength.

\subsection*{Acknowledgments}
We would like to thank G. Dvali and N. Weiner for useful discussions
and comments. Special thanks to A.E.C. Hern\'{a}ndez for providing
help with the Feynman diagrams.

\end{document}